\documentclass[12pt,preprint]{aastex}

\slugcomment{Submitted to: {\it The Astrophysical Journal}}

\shorttitle{CH at the Galactic Center}
\shortauthors{Magnani et al.}

\begin{document}

\title{CH 3 GHz Observations of the Galactic Center}

\author{Loris Magnani and Susan Zelenik}
\affil{Department of Physics and Astronomy, The University of Georgia,
    Athens, GA 30602}

\author {T. M. Dame}
\affil{Harvard-Smithsonian Center for Astrophysics, 
60 Garden St., MS 72, Cambridge, MA 02138}

\and 

\author {Ben Engebreth}
\affil{Colorado Center for Astrodynamics Research, University of Colorado at
Boulder, Boulder, CO 80309}

\begin{abstract}

A 3 $\times$ 3 map of the Galactic Center was made at 9$\arcmin$ resolution
and 10$\arcmin$ spacing
in the CH $^2\Pi_{1/2}$, J=1/2, F=1-1 transition at 3335 MHz.
The CH emission shows a velocity extent that is nearly that of the CO(1-0)
line, but the CH line profiles differ markedly from the CO.  The 3335 MHz
 CH transition primarily traces low-density molecular gas and our
observations indicate that the mass of this component within $\sim$ 30 pc of the 
Galactic Center is $\sim$ 9 $\times$ 10$^6$ M$_\odot$.
The CO-H$_2$ conversion factor obtained for the low-density gas in the
mapped region is greater than 
that thought to apply to the dense molecular gas at the Galactic Center.
In addition to tracing the low-density molecular gas 
at the Galactic Center, the CH spectra show evidence of emission from molecular 
clouds along the line of sight both in the foreground and background.
The scale height of these clouds ranges from 27 - 109 pc,
consistent with previous work based on observations of molecular clouds
in the inner Galaxy.
 
\end{abstract}

\keywords{Galaxy: center, ISM: molecules, radio lines: ISM}

\section{Introduction}

The Galactic Center (GC) is home to the greatest concentration of molecular
gas in the Galaxy.  The environment in the inner 300 pc (known
as the Central Molecular Zone or CMZ) contains molecular
clouds that have higher average densities and pressures than those in other regions 
($n \sim$ 10$^4$ - 10$^5$  cm$^{-3}$, P/$k \sim$ 10$^5$ K cm$^{-3}$  
- e.g., Blitz et al. 1993; Martin et al. 2004).
Most of the information on the dense molecular clouds in the CMZ has come 
from studies of the CO(1-0) transition and higher-density
molecular mass tracers (e.g., Jackson et al. 1996; 
Morris 1997; Tsuboi, Handa, \& Ukita 1999).  However, a survey of 
the C$^{18}$O(1-0) transition 
in the GC region by Dahmen et al. (1998), when compared to a similar CO(1-0)
survey by (Bitran et al. 1997), indicates that in addition to
the dense, hot, molecular clouds, there is a molecular gas component
with relatively low density ($\sim$ 10$^{2.5}$ cm$^{-3}$) and high kinetic temperature
($\sim$ 150 K)
whose total mass is a significant fraction of the better-studied denser component. 
A similar conclusion was reached by Oka et al. (1998) on the basis of CO(2-1)
observations of the GC.  These authors speculate that there are two components of
molecular emission associated with the molecular clouds at the GC: 
the well-established, high-density gas arises in 
molecular \lq\lq clumps" within the clouds 
that have relatively low filling factors, and a more 
pervasive \lq\lq diffuse" component has a filling factor of $\sim$ 1.  
Oka et al. (1998) associate the diffuse molecular component with individual molecular
clouds at the GC; for example, Sgr B2 is known to have a hot, low-density, molecular
envelope (H\"uttenmeister et al. 1995).  
However, the low-density molecular component may also
be produced by the strong tidal forces
at the GC that can shear clouds with densities less than a critical density
(e.g., Stark \& Blitz 1978; Stark \& Bania 1986).  In this case, the low-density
gas would likely fill the volume not occupied by the dense molecular clouds and  
also attain a surface filling factor of $\sim$ 1.
In order to survey this pervasive, diffuse, low-density 
molecular component,
we observed 0.25 square degrees in the direction of the GC in the
3335 MHz transition of methylidine (CH). 

The 3335 MHz CH line  
($^2\Pi_{1/2}$, J = 1/2 ground state, F = 1-1 main-line transition) is a good 
linear tracer of   
low-density molecular gas ($n <$ 10$^4$ cm$^{-3}$ -
Magnani et al. 2003 and references therein).
Because the transition is optically thin in the 
interstellar medium, many of the difficulties 
inherent in interpreting CO(1-0) data are avoided.
In addition to the molecular gas at the GC, an optically thin tracer
like the 3335 MHz CH line reveals more clearly the presence of
foreground and background gas along the entire line of sight.
In particular, all our CH spectra show a distinct, sharp, emission feature
at v$_{LSR} \sim$ 0 km s$^{-1}$. We argue in \S 4 that 
this feature arises from molecular
clouds in the near and far halves of the Galaxy whose systematic motion
is perpendicular to the line of sight. The width of the feature, in turn, 
reflects the radial random motions of this ensemble of clouds.  As
a measure of the one-dimensional random motions of the cloud ensemble,  
the line width can be used to determine the scale height of the molecular clouds
in the beam; something that cannot be done with optically thick transitions such
as the CO(1-0) line.  

In this paper, we present the most sensitive CH observations of the GC to 
date.  Unlike previous observations, the bandwidth 
of our observations is
sufficient to cover the full velocity extent of the 
Galactic Center CH emission.  In \S 2 we describe our
observations and review all previous CH observations of the GC in the
literature.  The
column densities of CH and H$_2$ and the molecular
mass traced by the CH 3335 MHz line are derived in \S 3. We discuss how
the CH emission compares to the CO(1-0) emission over 
similar velocity intervals and 
what the differences between the two tracers reveal about
the molecular gas distribution at the GC. 
The value of N(H$_2$) derived
from the CH data allows us to determine the CO-H$_2$ conversion factor
in the diffuse molecular gas, and we compare the conversion factor for
the diffuse and dense components. 
In \S 4 the scale height of molecular gas
along the line of sight is
determined and compared to previous work.  A short summary
closes the paper.

\section{Observations}

The CH 3335 line was observed in the direction of the GC during 1999 March
using the now-defunct  
NRAO\footnote{The National Radio Astronomy Observatory (NRAO) is operated
by Associated Universities, Inc., under contract with the National Science 
Foundation.} 140 ft telescope in Green Bank, West Virginia. 
At 3.3 GHz the beam size of the 140 ft was 9$\arcmin$ 
which, at the distance of the GC (taken to be 8 kpc for the
remainder of the paper), is 21 pc.  The observing
configuration consisted of a front end with a corrugated dual-hybrid
mode feed in which two linear polarizations were fed into a dual-channel
FET amplifier receiver.  The system temperature on the sky was in the
range 50 - 80 K, depending on the atmospheric conditions, the 
antenna elevation, and the continuum flux at 3 GHz.  
The autocorrelator was configured into two sections
of 512 channels, with each section covering a bandwidth of 10 MHz at a
velocity resolution of 1.8 km s$^{-1}$ per channel.  The total velocity
coverage of each spectrum was $\sim$ 900 km s$^{-1}$ centered on v$_{LSR} =$
0 km s$^{-1}$.

A 3 $\times$ 3 map of the GC region was made with the central spectrum 
at $\ell =$ 0$\arcdeg$, $b =$ 0$\arcdeg$ and all other spectra offset
by 0.125$\arcdeg$ in latitude and/or longitude. Each position was observed
in ON-OFF mode with one hour total on-source integrations.  The OFF positions
were determined from the catalog of Verter et al. (1983).
For each line of sight, 
the two polarizations were added together and the resulting spectrum was
baselined with a polynomial of
order 6 (discussed in the following section) and Hann smoothed
to yield an {\it rms} noise level of $\sim$ 5-8 mK per channel.
A raw spectrum for one of the lines of sight 
before baselining or Hann smoothing is shown in Figure 1.
Baseline fitting windows were determined by looking at
the raw data as in Figure 1 and by using CO spectra from 
Bitran et al. (1997) to indicate the maximum extent of
the molecular emission.
Individual reduced spectra for each position are shown in Figures 2$a$ - 2$i$ 
along with the corresponding CO(1-0) spectra from the Bitran et al. (1997) survey of
the GC.  The CO data are at comparable angular and velocity resolution 
(8.8$\arcmin$ and 1.3 km s$^{-1}$, respectively).

Given the historical dearth of radio telescopes with 3 GHz receivers, there are very
few observations of the GC in the CH ground state, hyperfine transitions at 3 GHz.
Moreover, previous observations of CH at the GC did not have sufficient
sensitivity and, often, bandwidth, to detect the broad component seen
in our spectra, and tended to focus on individual, narrow, emission features at the
GC (Gardner \& Robinson 1974; Gardner, Robinson, \&
Sinclair 1976; Whiteoak, Gardner, \& Sinclair 1978; Genzel et al. 
1979; Whiteoak et al. 1985). Thus, the extended, broad CH component described in
this paper has not been noted before.

\section{Results}

The CH spectra presented in Figures 2$a$ - 2$i$ 
 all show a velocity extent that is nearly that of the
CO(1-0) emission.
However, the CH line profiles look markedly different from the 
corresponding CO profiles.  This is in contrast to the results of
Magnani, Lugo, \& Dame (2005) who compared CH and CO for 15 lines
of sight along the Galactic plane,  In those instances, the CH and
CO line profiles are strikingly similar, indicating that most of 
the gas in these clouds is at low density.  Blitz (1991) quotes 
an average value of Galactic plane GMCs of 50 cm$^{-3}$; three 
orders of magnitude lower than for the clouds at the GC.  Given
the very different nature of the molecular clouds in the plane
vs. the GC, it is not surprising  that in the former case the 
CO and CH profiles are very similar, while in the latter they are
different.

However, some of the GC molecular gas {\it is} at low density:
The widespread molecular component reported by Dahmen et al. (1998)
has physical parameters ($n \sim$ 10$^{2.5}$; T $\sim$ 150 K) ideal 
for CH 3 GHz observations (however, see \S 3.4).  Following Dahmen et al. (1998)
we will refer to this molecular gas as the ``thin" component.
The CH emission evident in Figures 2$a$ -2$i$ is likely tracing
the thin gas and, at some level, the denser gas from GMCs in 
the region.  Unfortunately, without extensive CH mapping of
GMCs, both at the GC and elsewhere in the plane,
it is not possible to determine the fraction of CH emission
that arises from each component.  Even though 
the CO and CH line profiles are
different, it is not surprising that the velocity extent is similar; if
tidal stripping of molecular clouds produces the diffuse thin
gas, then it likely fills the CMZ, and its 
velocity extent should be considerable given the GC gravitational 
potential. In the next section we will examine the relationship 
between the CO and CH emission in detail.

\subsection{Comparison of velocity-integrated CO and CH}

Because of the large velocity extent of the CH emission, we compare 
in this section the CO and CH data over similar velocity intervals.
Given the complicated noncircular motions at the GC (e.g.,
Morris \& Serabyn 1996), gas at velocities separated by only a few
km s$^{-1}$ within our 9$\arcmin$ beam 
can arise from very different regions.
Thus, in order to compare emission from similar regions, we broke up
each CH and CO spectrum into a series of equivalent velocity intervals.
Each interval extends 9 - 10 km s$^{-1}$ in velocity and comprises
5 or 6 channels of the CH spectra and 7 or 8 channels of the CO spectra.
Despite the differing velocity resolutions, the velocity intervals were
matched as closely as possible and differ by no more than 1 km s$^{-1}$ at
either extreme.  In this manner, dozens of data points from each spectrum 
comparing the  
velocity-integrated antenna temperatures for both species [defined as
W$_{CH}$ and W$_{CO}$, respectively] can be analyzed.. 

The datasets at a given
$b$ for the 3 longitudes in our map are divided into positive,  negative,
and near 0 
LSR velocities.  The parameters of the best-fit line 
to the positive and negative data are calculated excluding the velocity  
interval closest to 0 km s$^{-1}$ (i.e., the interval [$-$9, 0  
 km s$^{-1}$] for the negative velocities and [0, 9 km s$^{-1}$]
for the positive velocities).  The results are shown in 
Table 1 and reproduced in graphical
form in Figure 3.

The CH emission centered on 
v$_{LSR} =$ 0 km s$^{-1}$ is composed of emission from the GC
and also from foreground or background molecular
gas with respect to the GC (see \S 4). The least squares fit to the 18 data points
in this set did not show any W$_{CH}$-W$_{CO}$ correlation.  A glance
at Figure 3 indicates that the data from this component clearly differs from
the positive and negative datasets.  Because the molecular gas in this 
dataset does not arise entirely at the GC, we will not discuss it further.

Breaking up the CH emission into $\sim$ 9 km s$^{-1}$ intervals allows for a 
meaningful comparison with translucent and dark cloud data.  Magnani \& Onello
(1995) and Magnani et al. (1998) observed CO and CH from 48 lines of sight in 
translucent clouds and 12 in dark clouds.  The slopes of the W$_{CH}$-W$_{CO}$  relation
for those two data sets (8.2 and 10.1, respectively) are virtually identical to 
the slopes of the relation for positive and negative v$_{LSR}$ total points in Table 1.

The results shown in Table 1 indicate that despite clear differences between
the CO and CH profiles shown in Figure 2, there is a general correlation between
the CO and CH emission over similar velocity intervals.  Moreover, the slope 
of the W$_{CH}$ - W$_{CO}$ relation is similar to that determined 
previously for a sample of local dark and translucent clouds.  
Similar slopes for local and GC clouds imply that 
the physical conditions responsible for CH 3335 MHz emission from 
the molecular gas at the GC are likely similar to those in local gas. 
This re-enforces our contention that the CH 3335 MHz emission from the GC arises
primarily in low-density molecular gas - just as is the case for CH emission
in local molecular clouds.

\subsection{N(H$_2$) from W$_{CH}$ and W$_{CO}$}

All the CH spectral profiles consist of a 
broad, velocity-extended component (more than 350 km s$^{-1}$
FWZP)  and a
distinct spike component at $\sim$ 0 km s$^{-1}$.
In \S 4 we argue that the spike feature arises from molecular gas
outside the GC region.  By determining the zeroth moment of
 the CH emission after
subtracting the contribution from the
0 km s$^{-1}$ component (determined by fitting a Gaussian to the
spike), we can derive the
column density of CH at the GC using the standard relationship between
W$_{CH}$ and N(CH) (Rydbeck et al. 1976).
The values of W$_{CH}$ and N(CH) for 
the broad component for all the observed positions are given in Table 2. 
The largest 
source of uncertainty in the analysis is produced by the baseline fit to the raw spectra. 
Varying the order of the polynomial used for the baseline from 4 to
8 produced variations in the integrated antenna temperature of the
broad component of up to 50\%, but did not change the velocity 
extent of the emission.  With the demise of the 140 ft 
telescope, there is no possibility, at the moment, of re-observing
the GC at 3 GHz from the Western Hemisphere in order to confirm
the values of N(CH) in Table 2.  Thus, the numbers we 
derive below are uncertain at the 50\% level because of baseline 
uncertainties. 

The relationship between N(CH) and N(H$_2$)
is linear for values of N(CH) less than 
2 $\times$ 10$^{14}$ cm$^{-2}$, corresponding to N(H$_2$) $\le$
5 $\times$ 10$^{21}$ cm$^{-2}$ (Mattila
1986; Rachford et al. 2002; Magnani et al. 2003; Weselak et al. 2004). 
Using the relationship between E(B-V) and total hydrogen column
density [Bohlin, Savage, \& Drake (1978)], and a value for R$_V$
of 3.1 (e.g., Sneden et al. 1978), a column density of 5 $\times$ 10$^{21}$ cm$^{-2}$
corresponds to a visual extinction of nearly 3 magnitudes, squarely
in the translucent molecular gas regime (van Dishoeck \& Black 1988).
However, for N(H$_2$) greater than 5 $\times$ 10$^{21}$ cm$^{-2}$,
the linearity of N(CH) and N(H$_2$) begins to break down.  This was evident even 
in the first large-scale surveys of CH (Rydbeck et al. 1976; Hjalmarson et al. 1977).
Recently, using a compendium of CH data including
observations from the FUSE satellite, Liszt and Lucas (2002) also note a marked 
decline in the CH/H$_2$ ratio as N(H$_2$) increases from diffuse to 
dark cloud values. Moreover, theoretical chemical models of molecular clouds
invariably show that the CH abundance decreases rapidly at high extinctions or
H$_2$ volume densities (e.g., Viala 1986; Lee, Bettens, \& Herbst 1996).

The most comprehensive empirical study of the N(CH)-N(H$_2$) 
relation was conducted by Mattila (1986).
Figure 10 of his paper shows a marked deviation from linearity
at values of N(H$_2$) $>$ 10$^{22}$ cm$^{-2}$.  However, this deviation is based on 
only 6 data points from CH observations of 
GMCs including two lines of sight to Sgr A and
Sgr B2 in the GC. The CH data for Sgr A and Sgr B2 are taken from Genzel et al.
(1979) and, as mentioned in \S 2, do not have sufficient velocity coverage
or sensitivity to reveal the CH emission in its entirety.  Thus, the values of N(CH) 
quoted by Mattila (1986) for
at least those two points are underestimated and should be considered lower
limits to N(CH).

The question of how significantly the CH 3335 MHz line underestimates N(H$_2$) in 
GMCs has been addressed by  
Magnani, Lugo,  \& Dame (2005) who examined the relationship between CH and H$_2$
for a small sample of lines of sight through GMCs 
along the Galactic plane.  They demonstrate that, for 10 lines 
of sight clustered at ($\ell, b$) = (50$\arcdeg$, 0$\arcdeg$) and (110$\arcdeg$, 0$\arcdeg$),  
the 3335 MHz line underestimates N(H$_2$) by only a factor of
2-3.  This result is likely a consequence that a large fraction of a GMC's volume is
composed of low-density gas (e.g., Blitz 1991; Lada, Bally, \& Stark 1991). 

As mentioned above, Dahmen et al. (1998) and Oka et al. (1998) have
shown that not all the molecular gas at the GC is at high-density. 
Thus, the CH 3335 MHz line can still effectively trace some of the molecular
gas. Dahmen et al. (1998) estimate that $\sim$ 1.0 $\times$ 10$^7$ M$_\odot$
is in the thin gas regime, while the total molecular gas mass ranges from 
3.1 - 7.0 $\times$ 10$^7$ M$_\odot$  (Sodroski et al. 1994; Blitz et al. 
1985).  However, these estimates are for the central 600 pc of the GC; a 
much larger volume than that covered by our observations. It cannot be assumed
that the ratio of thin to dense gas remains constant as one nears the GC. 
Without knowing
the fraction of thin molecular gas in the area covered by our
observations, we cannot estimate the total amount of N(H$_2$) in the region
solely on the basis of the CH data; but we can determine N(H$_2$) in the
thin gas [defined as N(H$_2$)$_{\rm thin}$]. 
Table 2 shows that N(H$_2$)$_{\rm thin}$ derived from
the CH observations ranges from 5.3 $\times$ 10$^{22}$ to 1.5 $\times$
10$^{23}$ cm$^{-2}$ with an average of 9.6 $\times$ 10$^{22}$ cm$^{-2}$. 

Using a distance of 8 kpc for the GC, 0.25 square degrees as the size of
the observed region, and the average N(H$_2$) derived above, we obtain a lower 
limit for the thin molecular gas in the mapped region of 9 $\times$ 10$^6$ M$_\odot$.

\subsection{X$_{CO}$ at the Galactic Center}

In order to obtain N(H$_2$) from CO(1-0) data, an empirically-derived  
CO-H$_2$ conversion factor - defined as X$_{CO} =$ N(H$_2$)/W$_{CO}$, where
W$_{CO}$ is the velocity-integrated CO(1-0) antenna temperature - is used.  
Typical  values
of X$_{CO}$ in the Galactic plane range from 1.6
 - 4 $\times$ 10$^{20}$ cm$^{-2}$
[K km s$^{-1}$]$^{-1}$ (e.g., Combes 1991; Strong \& Mattox 1996;
Hunter et al. 1997; Dame, Hartmann, \& Thaddeus 2001 - 
 we drop the units of X$_{CO}$
for the remainder of the paper for brevity). If we use a value of 
1.8 $\times$ 10$^{20}$ as derived from far-infrared calibration
mainly from the solar neighborhood by Dame, Hartmann, \& Thaddeus
(2001), N(H$_2$) for the 9 CO
spectra shown in Figures 2$a$ - 2$i$ ranges from 0.89 - 2.7 x 10$^{23}$ cm$^{-2}$. 
However, there is strong evidence that X$_{CO}$ at the GC is probably lower 
than the Galactic value.  

Blitz et al. (1985) proposed a lower X$_{CO}$ at the GC based on a deficit of
gamma-rays in the region. Later,
Sodroski et al. (1994) suggested, based on dust-to-gas ratio arguments,
that the value of X$_{CO}$ in the GC region is lower by a factor of 4-9
than the disk value.
Similar results were found by Oka et al. (1998) and Sakano et al.
(1999).  Thus, instead of 1.8 $\times$ 10$^{20}$, X$_{CO}$ at the GC
is more likely in the 0.2 - 0.5 $\times$ 10$^{20}$ range. 

Magnani \& Onello (1995) describe in detail how the CH 3335 MHz
transition can be used to determine X$_{CO}$ in translucent 
molecular clouds.  If we apply this technique to the CH data
presented here to determine X$_{CO_{\rm thin}}$, 
a value of 0.8 $\times$ 10$^{20}$ is obtained.  This is
a lower limit because even if all the CH emission comes from
N(H$_2$)$_{\rm thin}$, the CO(1-0) emission still arises from 
both the thin and the dense molecular components.  If 
W$_{CO_{\rm dense}}$/W$_{CO_{\rm thin}}$ is proportional to the
ratio of the mass in the thin gas to that in the dense gas, then
W$_{CO_{\rm dense}}$/W$_{CO_{\rm thin}}$ ranges from 3-7 (see \S 3.2).
In turn, X$_{CO_{\rm thin}}$/X$_{CO_{\rm dense}}$ would range over the
same values. 

The CH data coupled with the above argument indicate that X$_{CO_{\rm thin}}$ may be in the 
2 - 6 $\times$ 10$^{20}$ range.
This result does not necessarily contradict that of
Dahmen et al. (1998) who find that the lower values of X$_{CO}$ proposed
by Sodroski et al. (1994) also provide good agreement between the CO(1-0)
GC mass and their estimate based on C$^{18}$O(1-0)
mapping of the region.  The molecular component they are referring to
is the dense component.  Although they derive a mass for the thin component,
they do not determine what value of X$_{CO}$ might be appropriate for it.
However, they do point out that the
molecular gas seen in the CO(1-0) line but not in the C$^{18}$O transition
is probably not virialized and the CO emission for this 
component may not be optically thick, in
contrast with the dense gas in the GC GMCs.
Given such disparate physical conditions for the various components
of the molecular gas, usage of a single coversion factor for all the GC molecular gas is
likely not valid.  It may be that the best way to determine X$_{CO}$
empirically for the thin gas is by using the CH method.  The first step to
address this issue would be to determine the complete extent of the 
CH emission from the GC region, and then to make a detailed comparison with
the Bitran et al. (1997) and Dahmen et al. (1997) data. 
In the section below, we discuss this type of comparison for the limited
region we mapped.

\subsection{Comparison of the CH 3335 MHz and C$^{18}$O transitions}

The C$^{18}$O data used by Dahmen et al. (1998) to argue for a diffuse,
warm molecular component at the GC is presented by Dahmen et al. (1997).
The spectra were taken with the Southern Millimeter-Wave Telescope - just like the
CO(1-0) data described above - and thus have 
similar spatial and velocity resolution to the CH data.  The sampling of
the C$^{18}$O observations is on a slightly coarser grid than our data
(0.15$\arcdeg$ vs. 0.125$\arcdeg$), but the difference is small enough
that we can make a direct comparison between our 9 CH spectra and the
9 C$^{18}$O spectra taken by Dahmen et al. and 
centered on and around ($\ell, b$) = (0$\arcdeg$,
0$\arcdeg$).

It is immediately clear that the velocity of the CH emission is
substantially more extended than that of C$^{18}$O.  For instance,
the C$^{18}$O spectrum at ($\ell, b$) = (0$\arcdeg$, $-$0.125$\arcdeg$) shows
emission only from $-$70 to 100 km s$^{-1}, $\footnote{The emission near
$-$200 km s$^{-1}$ is produced by the HNCO(5$_{05}$ - 4$_{04}$) transition.}
while the CH 3335 MHz emission clearly extends from $-$175 to 200 km
s$^{-1}$. This behavior is similar for all nine CH lines of sight.
Because the C$^{18}$O directly traces the dense
molecular clumps in the GMCs at the GC, it is the {\it absence} of
C$^{18}$O(1-0) emission compared to
CO(1-0) emission that led Dahmen et al. (1998) to conclude, on the 
basis of LVG models, that the C$^{18}$O-deficient regions likely
contained warmer, lower density, molecular gas.

The CH 3335 MHz emission tracks the CO(1-0) emission in velocity 
very well, and indicates that the thin gas component
arises predominantly in those regions that produce the most extreme
velocities of molecular emission.  This behavior is consistent with
what would be expected from a molecular component produced by tidal
stripping of gas from GMCs at the GC; this component would fill 
the region and rapidly assume a velocity distribution commensurate
with the GC potential.  In order to confirm this idea, more extensive
CH mapping of the GC region should be done.  At the moment, only
the Parkes radiotelescope in Australia is equipped for this endeavor.

\section{The Molecular Scale Height of the Galaxy}

All the CH profiles show a narrow emission feature at v$_{LSR}$
 $\sim$ 0 km s$^{-1}$ .
The CO data show this feature clearly only 
in the spectra at $b = +$0.125$\arcdeg$.
The other spectra do not show a spike at this velocity and may even have
evidence of self-absorption (cf. the spectra at $b = -$0.125$\arcdeg$).
This very different behavior of the CH 3335 MHz and CO 115 GHz line is
most likely due to the very different opacities of the two transitions.  The
optically thin CH line is picking up emission from 
all the clouds along the line of sight in both the
near and far halves of the Galaxy, while the CO emission
at $\sim$ 0 km s$^{-1}$ from the GC is 
opaque and dominates the spectral profiles at that velocity.  
The CO spectra at $b = +$0.125$\arcdeg$
show less overall CO  emission 
than the others so the emission $\sim$ 0 km s$^{-1}$ 
is not overwhelmed and is easier to discern. 
Figure 4 shows 
the composite CH spectrum of the mapped region, and 
Table 3 shows the parameters of the Gaussian fits to both the
composite and individual spectral profiles.

The optically thin CH spike at 0 km s$^{-1}$ is likely sampling the 
emission from molecular clouds in the foreground and background with
respect to the GC whose 
motion is primarily transverse to
the line of sight. Thus, the width of the narrow feature samples the radial  
one-dimensional velocity dispersion of the cloud ensemble in the beam.
A simple numerical simulation shows that about 2/3 of this emission arises
from clouds in the near half of the Galaxy while the remainder of the 
emission comes from clouds beyond the GC. The simulation
populates a solid angle the size of the CH beam with equivalent CH-emitting
units, representing GMCs, at varying distances from the Sun. The number 
of units in the beam is an input parameter. Beyond a distance of 6 kpc,
beam dilution decreases the contribution of each CH-emitting unit (this is
equivalent to assuming that the CH-emitting units represent GMCs about 15 pc
in diameter) by a factor
of (6/d)$^2$, where d is the distance of the unit from the Sun in kpc.

Using the relation given by Magnani et al. (2000), we can calculate the
scale height of the clouds in the beam given the one-dimensional 
velocity dispersion of the clouds, the stellar 
scale height, and  mass surface density in the Inner Galaxy.
The velocity dispersion of the  
clouds is readily obtained from the FWHM of the
Gaussian fits to the CH narrow feature, the mass surface density
is taken to be 50 $\pm$ 10 M$_\odot$ pc$^{-2}$ (Kuijken \& Gilmore
1991; Flynn \& Fuchs 1994), and the stellar scale 
height is 300 $\pm$ 20 pc (Gilmore \& Reid 1983; Binney \& Merrifield
1998).  With
the preceding values, the molecular scale heights for 
the individual lines of sight vary between 27 and 73 pc (see Table 2),
and for the composite profile the scale height is 109 pc.  These values
are similar to the values obtained from CO observations
of GMCs (88 pc - Fich \& Blitz 1984; 65-80 pc - Scoville \& Sanders 1987; 
51 pc - Bronfman et al. 1988; 74 pc - Dame et al. 1987; 35 pc - Stark and Lee
2005). The general
agreement of the molecular scale height derived here with that derived
from CO implies that the bulk of the molecular gas in the Galactic disk
is moving on very nearly circular orbits.  We do note that
the scale height in the composite profile may indicate a slightly
larger scale height for the molecular gas, 
more reminiscent of that of the local,
small molecular clouds (e.g., Magnani, Blitz, \& Mundy 1985).
It would be useful to probe a larger region to study the 
variation in linewidth of this feature as a function of position.

\section{Summary}

We have presented the most sensitive and velocity-extended 
 CH 3335 MHz observations of the GC to date.  The CH
emission profiles cover nearly the same velocity extent as
CO spectra of the corresponding regions, though the shapes of the 
profiles are markedly different.   The values of N(H$_2$)
at the GC obtained from the CH data range from 5.3 $\times$
10$^{22}$ - 1.5 $\times$ 10$^{23}$ cm$^{-2}$. 
The CH emission is likely produced by a low-density, intercloud,
molecular component which pervades the GC, and a component associated
with the outer envelopes of GMCs at the GC.  The relative 
contribution from each source to the CH profile is yet to be determined.

The CO-H$_2$ conversion factor, X$_{CO}$, can be determined from
the CH data for the lower density molecular component described
above.  The resulting value, 0.8 $\times$ 10$^{20}$, is lower than
the values obtained for disk GMCs but is likely underestimated by 
a factor 3 - 7.  This implies that X$_{CO}$ is greater for the 
lower density gas than for dense GC GMCs (whose X$_{CO}$ is
thought to be in the 0.2 - 0.5 $\times$ 10$^{20}$ range). 

The mass of molecular gas within $\sim$ 30 pc of the GC as determined 
from the CH data is $\sim$ 9 $\times$ 10$^6$ M$_\odot$. Although
the mapped region was fairly small (30$\arcmin \times$ 30$\arcmin$),
the CH 3335 MHz emission was readily detected for all the observed
lines of sight.
 A more complete survey of the GC in the CH $^2\Pi_{1/2}$, J=1/2,
F=0-1, 1-1, and 1-0 transitions may better trace the
lower-density molecular gas than conventional CO
surveys and elucidate the relation between
this diffuse molecular component and atomic hydrogen at
and around the GC. 

An unexpected consequence of the CH survey of the
GC was the detection
of prominent emission
at v$_{LSR} \sim$ 0 km s$^{-1}$. This feature  most likely arises
from foreground and background clouds with respect to the GC and
can be used to determine the scale height of this ensemble.
The values we obtain (27 - 109 pc) are similar to the scale height of
GMCs in the Inner Galaxy as determined from CO surveys.

\acknowledgments

Part of the work here was undertaken while S.Z. was a summer intern at
the University of Georgia under a Research Experience for Undergraduates
program sponsored by the National Science Foundation (PHY 00-97457).
We thank G\"oran Sandell for a critical reading of an early version of
the manuscript.  We also thank an anonymous referee for comments that  
greatly improved the presentation of the results and the organization of
the paper; in particular, with regard to the section on X$_{CO}$.

\clearpage

\begin{deluxetable}{rrrrr}
\tabletypesize{\scriptsize}
\tablecaption{CH - CO Intensity Relations - Linear Least Squares 
Fitting\tablenotemark{a}}
\tablewidth{0pt}
\tablehead{
\colhead{Dataset} & \colhead{Number of}   & \colhead{Intercept} &
\colhead{Slope} & \colhead{Correlation} \\ 
\colhead{}   & \colhead{points} & \colhead{} &
\colhead{$\times$ 10$^{-3}$} & \colhead{Coefficient} 
}
\startdata
 positive v$_{LSR}$\tablenotemark{b}         &     &           &         &       \\
 $\ell$ =    0.125$\arcdeg$ &  60 &    0.163  &  6.53   & 0.69 \\
 $\ell$ =    0.000$\arcdeg$ &  58 & $-$0.026  &  9.30   & 0.88 \\
 $\ell$ = $-$0.125$\arcdeg$ &  41 & $-$0.002  &  5.83   & 0.88 \\
 total                      & 159 & $-$0.011  &  8.19   & 0.83 \\
                                             &     &           &         &       \\
negative v$_{LSR}$\tablenotemark{c}          &     &           &         &       \\
$\ell$ =    0.125$\arcdeg$ &  51 &    0.028  & 22.21   & 0.89 \\
$\ell$ =    0.000$\arcdeg$ &  43 &    0.327  &  8.36   & 0.53 \\
$\ell$ = $-$0.125$\arcdeg$ &  48 &    0.105  &  3.95   & 0.34 \\
total                      & 142 &    0.087  &  9.95   & 0.56 \\
                                             &     &           &         &       \\
non-GC gas$\tablenotemark{d}$                &     &           &         &       \\
total                      & 18  &    1.29   &  -1.85  & -0.06 \\
\enddata

\tablenotetext{a}{In the form W$_{CH}$ = A + B W$_{CO}$, where A is the
y-intercept and B is the slope.}

\tablenotetext{b}{Excluding data points 
from velocity interval 0 $<$ v$_{LSR} <$ 9 km s$^{-1}$.  See \S 3.1 for details.}

\tablenotetext{c}{Excluding data points
from velocity interval $-$9 $<$ v$_{LSR} <$ 0 km s$^{-1}$.  See \S 3.1 for details.}

\tablenotetext{d}{Includes all data points from velocity interval 
$-$9 $<$ v$_{LSR} < +$9 km s$^{-1}$.  See \S 3.1 for details.}

\end{deluxetable}

\begin{deluxetable}{rrrrrrrrr}
\tabletypesize{\scriptsize}
\tablecaption{CH and CO Observations and Derived Quantities for the GC}
\tablewidth{0pt}
\tablehead{
\colhead{$\ell$ } & \colhead{$b$}   & \colhead{W$_{CH}$\tablenotemark{a}} 
& \colhead{W$_{CH}$\tablenotemark{b}} & 
\colhead{N(CH)\tablenotemark{c}} & \colhead{N(H$_2$)$_{\rm thin}$\tablenotemark{d}} & 
\colhead{W$_{CO}$\tablenotemark{e}} & \colhead{N(H$_2$)\tablenotemark{f}} &
\colhead{N(H$_2$)$_{CO}$/N(H$_2$)$_{CH}$ \tablenotemark{g}} \\
\colhead{degrees}   & \colhead{degrees} & \colhead{ K km s$^{-1}$} &
\colhead{ K km s$^{-1}$} &
\colhead{ cm$^{-2}$} & \colhead{ cm$^{-2}$} & \colhead{ K km s$^{-1}$} &
\colhead{ cm$^{-2}$}  &          
}
\startdata
 0.125 & 0.125  &  8.5 &  7.5 & 2.7 $\times$ 10$^{15}$  &  5.7 $\times$ 10$^{22}$ 
                &    888.8 &  1.60 $\times$ 10$^{23}$ &  2.8 \\
 0.125 & 0.000  & 19.2 & 18.3 & 6.5 $\times$ 10$^{15}$  & 13.7 $\times$ 10$^{22}$ 
                &   1515.7 &  2.73 $\times$ 10$^{23}$ &  2.0 \\
 0.125 & -0.125 & 16.5 & 15.3 & 5.4 $\times$ 10$^{15}$  & 11.4 $\times$ 10$^{22}$ 
                &   1278.0 &  2.30 $\times$ 10$^{23}$ &  2.0 \\
 0.000 & 0.125  &  9.8 &  8.5 & 3.0 $\times$ 10$^{15}$  &  6.3 $\times$ 10$^{22}$ 
                &    853.1 &  1.54 $\times$ 10$^{23}$ &  2.4 \\
 0.000 & 0.000  & 22.7 & 20.4 & 7.3 $\times$ 10$^{15}$  & 15.4 $\times$ 10$^{22}$ 
                &   1472.9 &  2.65 $\times$ 10$^{23}$ &  1.7 \\
 0.000 & -0.125 & 18.0 & 16.0 & 5.7 $\times$ 10$^{15}$  & 12.0 $\times$ 10$^{22}$ 
                &   1032.6 &  1.86 $\times$ 10$^{23}$ &  1.6 \\
-0.125 & 0.125  &  8.3 &  6.9 & 2.5 $\times$ 10$^{15}$  &  5.3 $\times$ 10$^{22}$ 
                &    494.0 &  8.89 $\times$ 10$^{22}$ &  1.7 \\ 
-0.125 & 0.000  &  9.6 &  8.0 & 2.9 $\times$ 10$^{15}$  &  6.1 $\times$ 10$^{22}$ 
                &   1171.3 &  2.11 $\times$ 10$^{23}$ &  3.5 \\
-0.125 & -0.125 & 15.3 & 14.0 & 5.0 $\times$ 10$^{15}$  & 10.5 $\times$ 10$^{22}$ 
                &   1237.2 &  2.23 $\times$ 10$^{23}$ &  2.1 \\
\enddata


\tablenotetext{a}{Velocity-integrated CH 3335 MHz antenna temperature. 
The uncertainty in this quantity is driven overwhelmingly
by the baseline fit (see \S 3.2}

\tablenotetext{b}{Velocity-integrated CH 3335 MHz antenna temperature for
the broad, extended component only (see \S 3).     
}

\tablenotetext{c}{N(CH) is derived from the integrated antenna temperature
in column 3 after correcting for the beam efficiency, the beam filling 
fraction, and assuming $\vert$ T$_{ex} \vert \gg$ T$_{bg}$. See Magnani \& Onello
(1995) for details.}

\tablenotetext{d}{N(H$_2$) derived from N(CH) via the relation established
by Mattila (1986). We refer to this gas as ``thin" for reasons elaborated in
\S 3.2}

\tablenotetext{e}{Integrated CO(1-0) line emission from the data of Bitran
et al. (1997).}

\tablenotetext{f}{N(H$_2$) derived from W$_{CO}$ using a conversion factor of
1.8 $\times$ 10$^{20}$.}

\tablenotetext{g}{Ratio of N(H$_2$) derived from CO(1-0) data divided by N(H$_2$)
derived from CH data.}
\end{deluxetable}

\clearpage

\begin{deluxetable}{rrrrrr}
\tabletypesize{\scriptsize}
\tablecaption{The CH Narrow Component at v$_{LSR} \sim$ 0 km s$^{-1}$}
\tablewidth{0pt}
\tablehead{
\colhead{$\ell$ } & \colhead{$b$}   & \colhead{T$_A$} &
\colhead{$\Delta$v} & \colhead{v$_{LSR}$} & 
\colhead{Scale Height \tablenotemark{a}}                    \\
\colhead{degrees}   & \colhead{degrees} & \colhead{mK } &
\colhead{km s$^{-1}$} & \colhead{km s$^{-1}$} & \colhead{pc}  
}
\startdata
 0.125 &  0.125 & 115  &  8.41   & $-$0.31  & 54 \\
 0.125 &  0.000 & 213  &  4.26   & $+$1.66  & 27 \\
 0.125 & -0.125 & 104  & 11.38   & $-$3.08  & 73 \\
 0.000 &  0.125  & 230  &  5.40   & $-$0.13  & 34 \\
 0.000 &  0.000 & 270  &  7.76   & $-$0.25  & 49 \\
 0.000 & -0.125 & 167  & 11.34   & $-$4.46  & 72 \\
-0.125 &  0.125 & 186  &  7.05   & $-$1.76  & 45 \\
-0.125 &  0.000 & 137  & 10.54   & $-$0.49  & 67 \\
-0.125 & -0.125 & 111  & 10.62   & $-$3.40  & 68 \\
 composite \tablenotemark{b}   &        & 182 & 17.61 & $-$1.20 & 109 \\
\enddata

\tablenotetext{a}{Scale height is derived using formulation given by
Magnani et al.  (2000 - see \S 4 for details).}

\tablenotetext{b}{Average of the 9 individual spectra.  See Figure 4.}

\end{deluxetable}

\clearpage

\begin{figure}
\caption{
Raw CH spectrum at one of the 9 positions ($\ell =$ 0.000$\arcdeg$, $b =$ 0.125$\arcdeg$).  The
spectrum is a one-hour on source integration on in ON-OFF mode, with both polarizations added together.
Emission is present from at least $-$100 to $+$180 km s$^{-1}$.
See \S 2 for more details.  The same data are presented after Hann smoothing and removal of a sixth order
baseline in the lower panel of Figure 2d.
}
\plotone{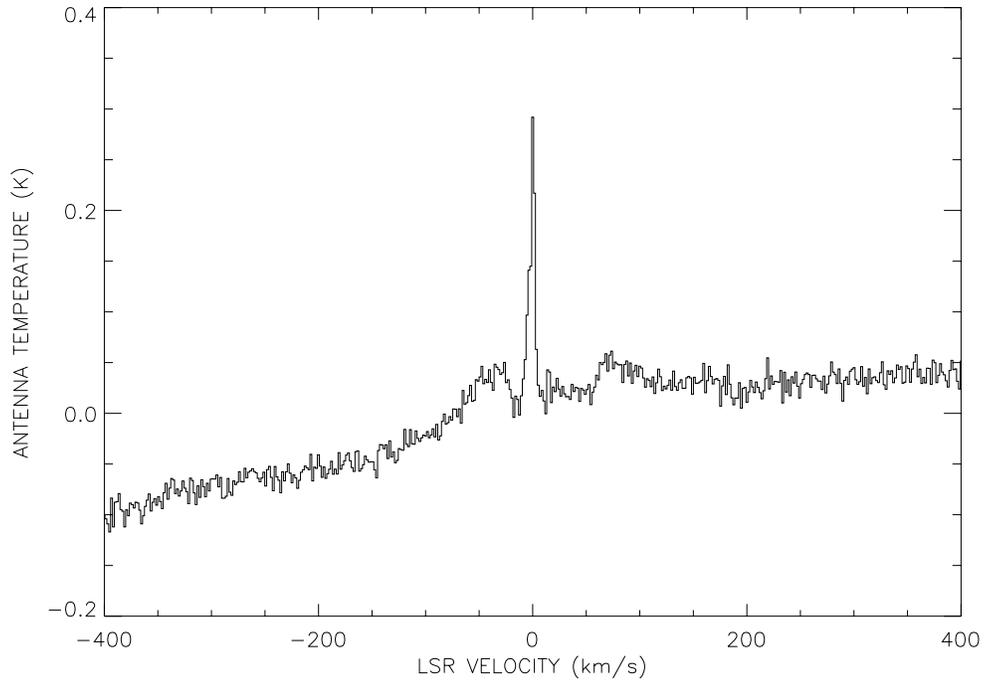}
\end{figure}

\begin{figure}
\figurenum{2a}
\caption{
CH and CO spectra for each of the nine observed lines of sight.
At bottom, spectrum of the CH $^2\Pi_{1/2}$ F=(1-1) transition 
at 3335 MHz for position $\ell =$
0.125$\arcdeg$, $b =$ 0.125$\arcdeg$. 
The beamsize is 9$\arcmin$ and the velocity 
resolution is 3.6 km s$^{-1}$ after Hann smoothing. 
Above, CO(1-0) spectrum
from Bitran et al. (1997) for the same position
 with a beamsize of 8.8$\arcmin$ and a velocity resolution of 1.3 km s$^{-1}$. 
See \S 2 for more details. 
}
\plotone{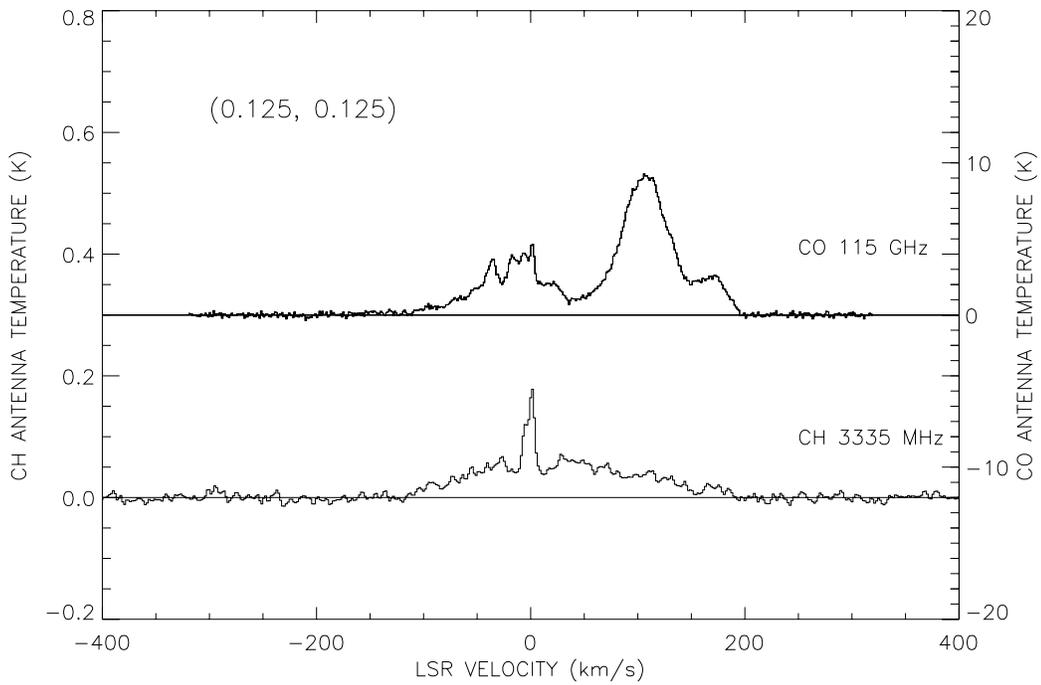}
\end{figure}

\begin{figure}
\figurenum{2b}
\plotone{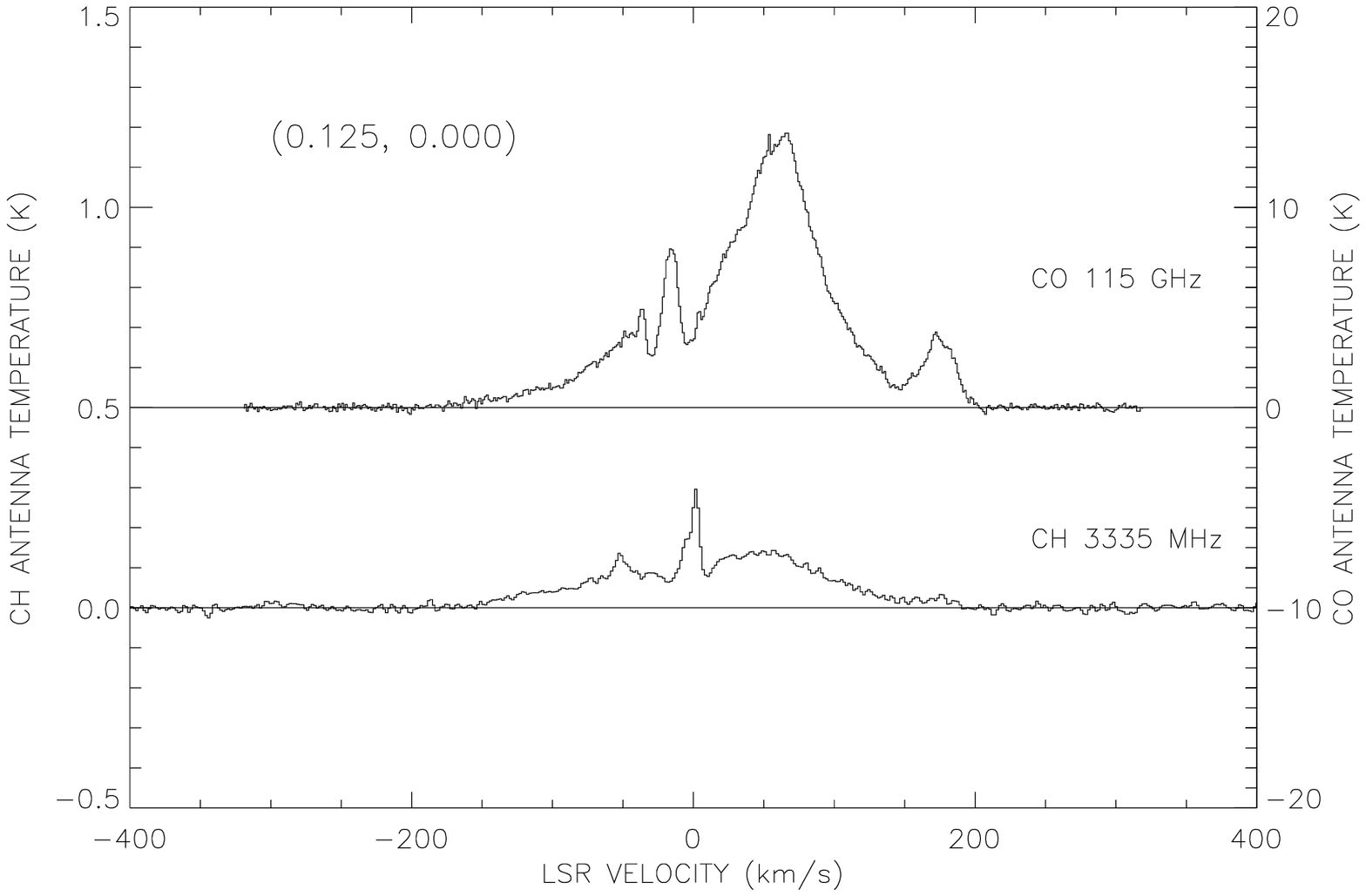}
\caption{
}
\end{figure}

\begin{figure}
\figurenum{2c}
\plotone{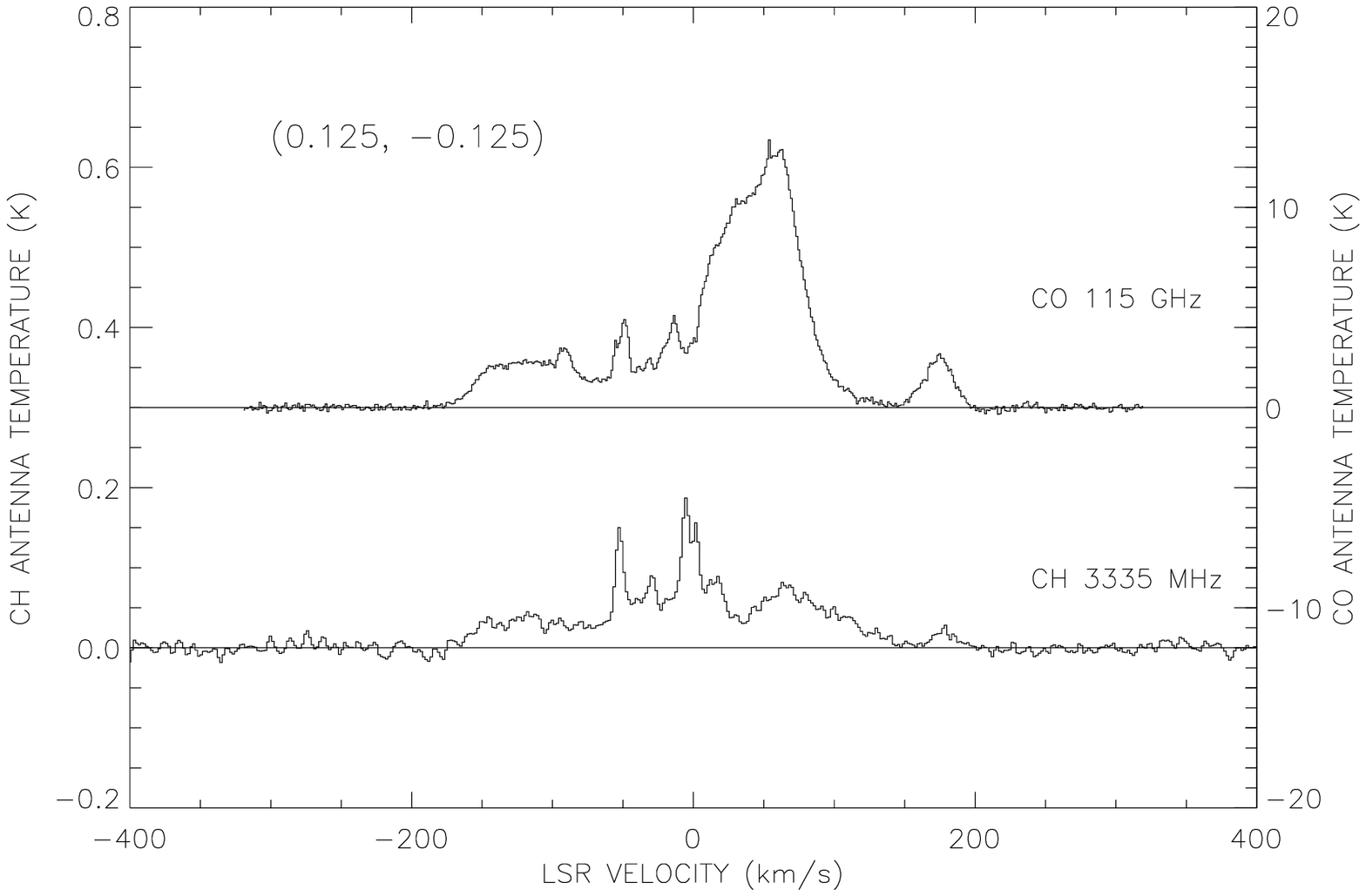}
\caption{
}
\end{figure}

\begin{figure}
\figurenum{2d}
\plotone{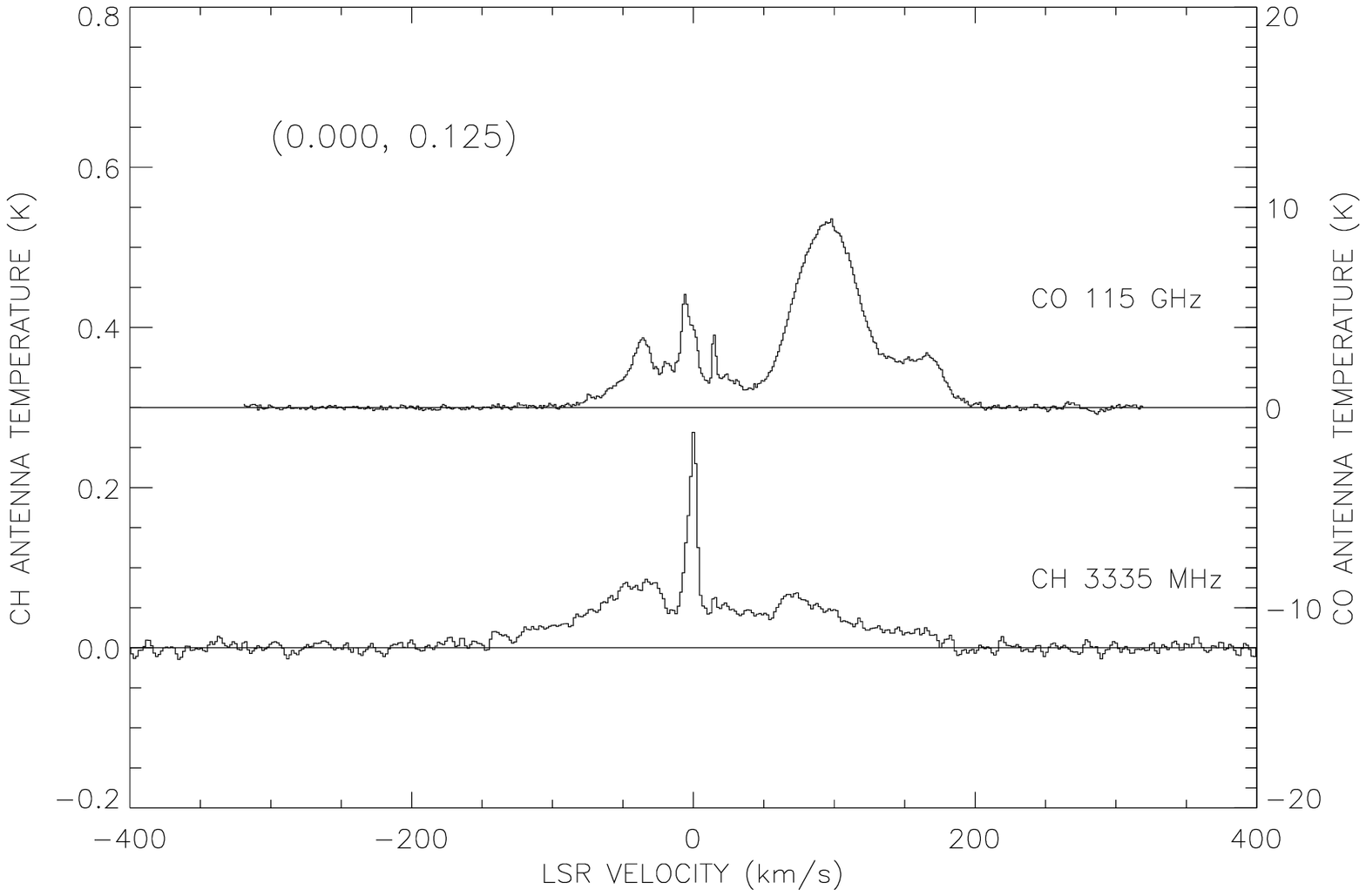}
\caption{
}
\end{figure}

\begin{figure}
\figurenum{2e}
\plotone{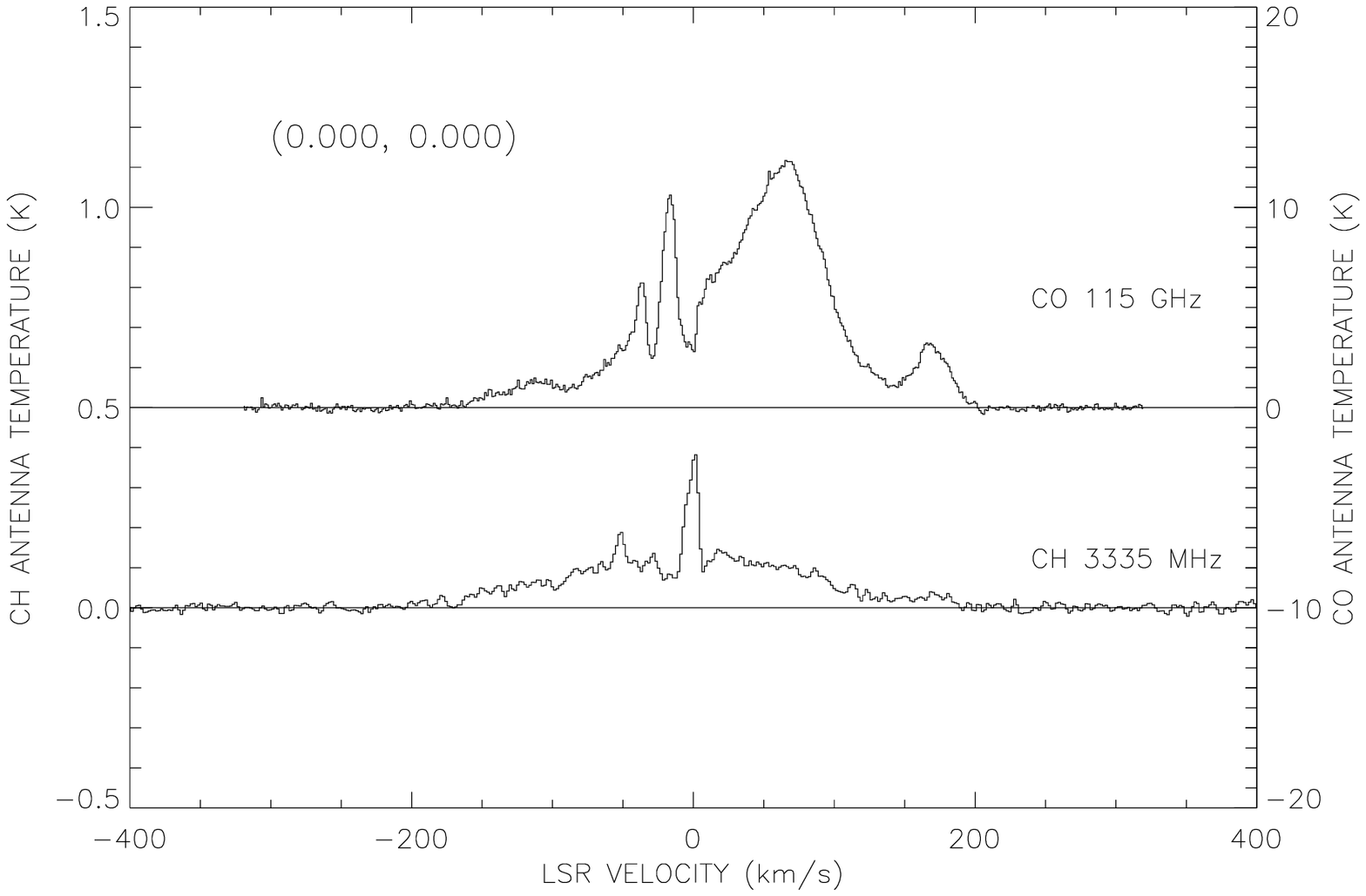}
\caption{
}
\end{figure}

\begin{figure}
\figurenum{2f}
\plotone{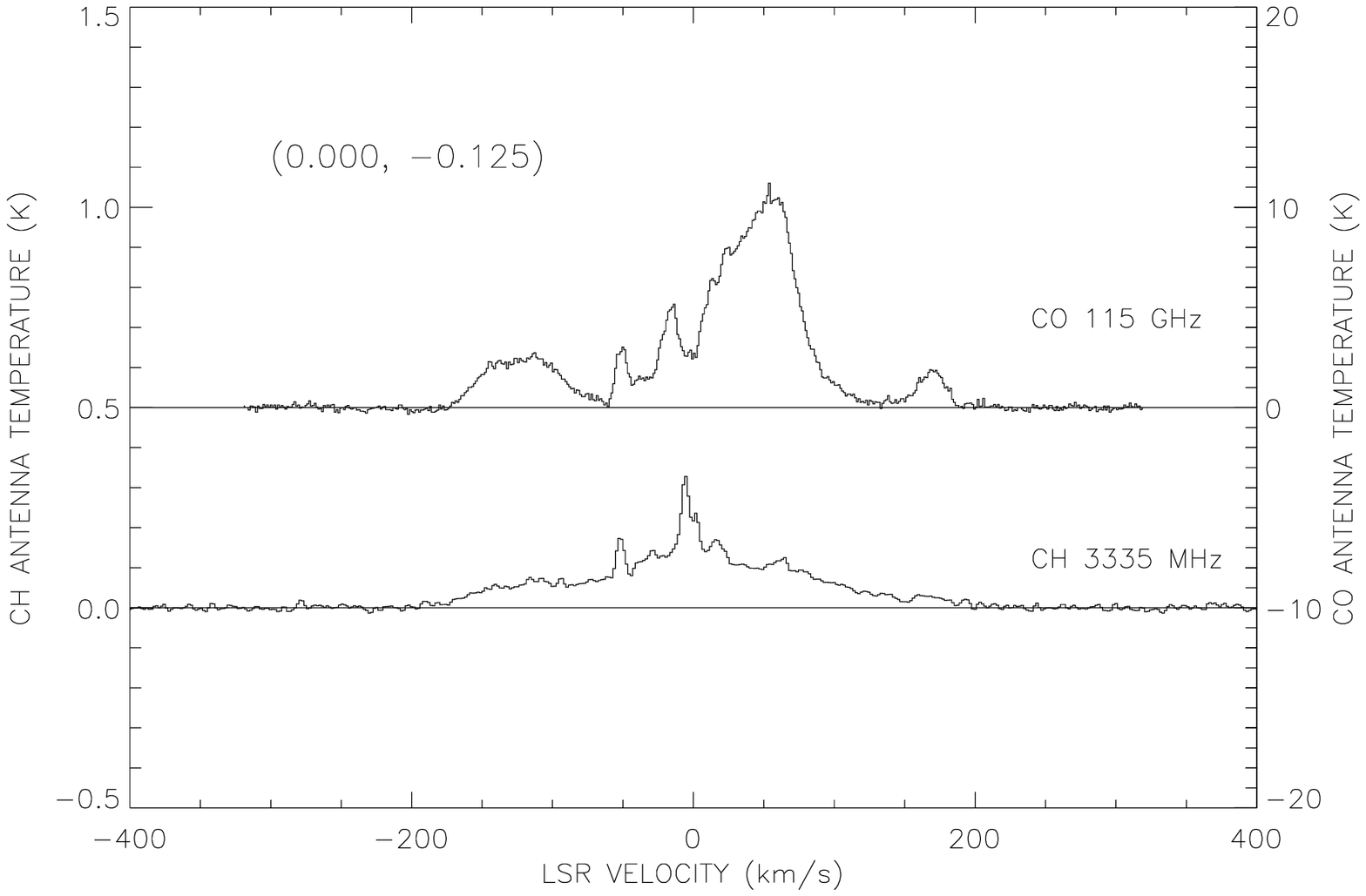}
\caption{
}
\end{figure}

\begin{figure}
\figurenum{2g}
\plotone{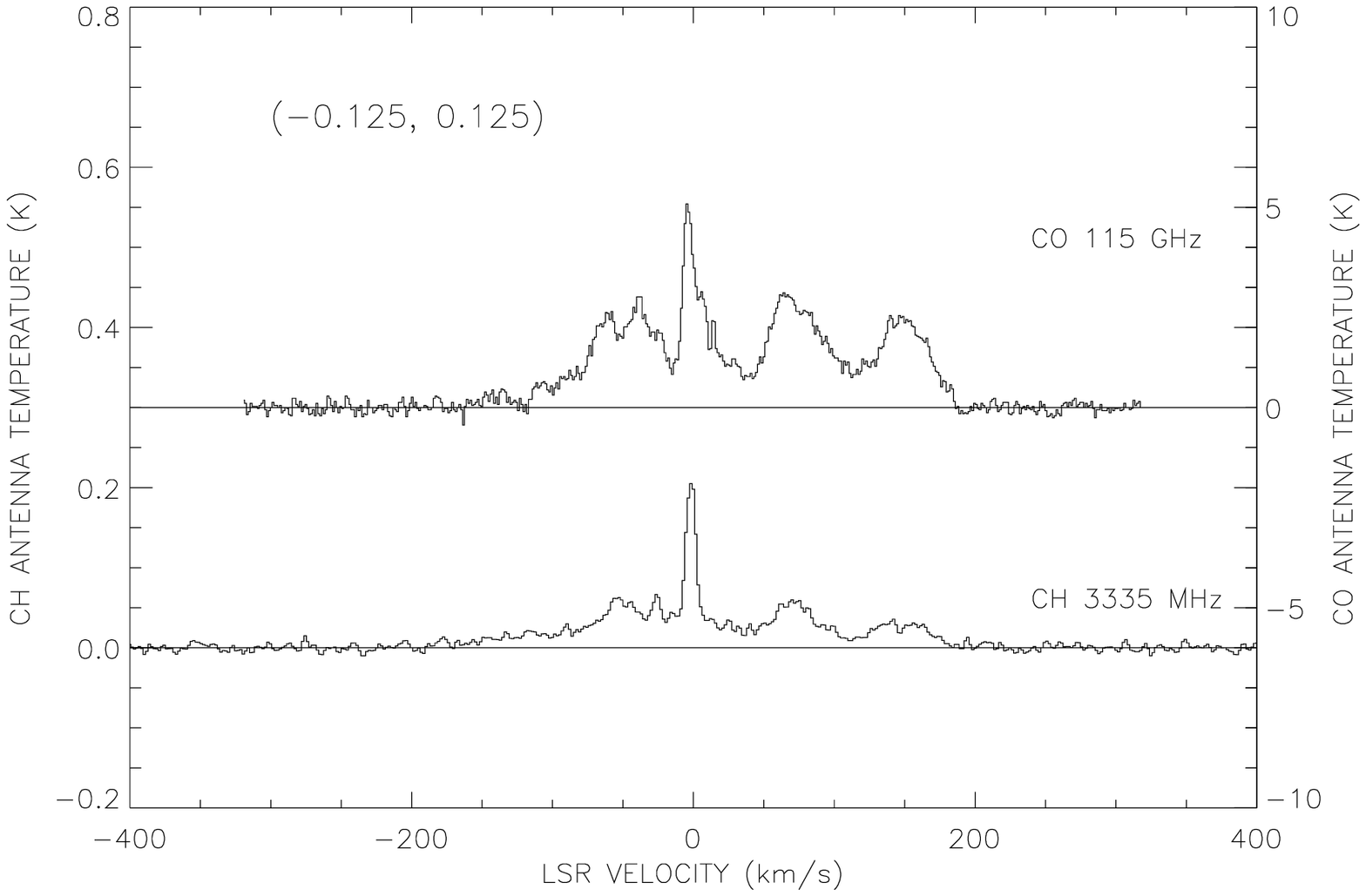}
\caption{
} 
\end{figure}

\begin{figure}
\figurenum{2h}
\plotone{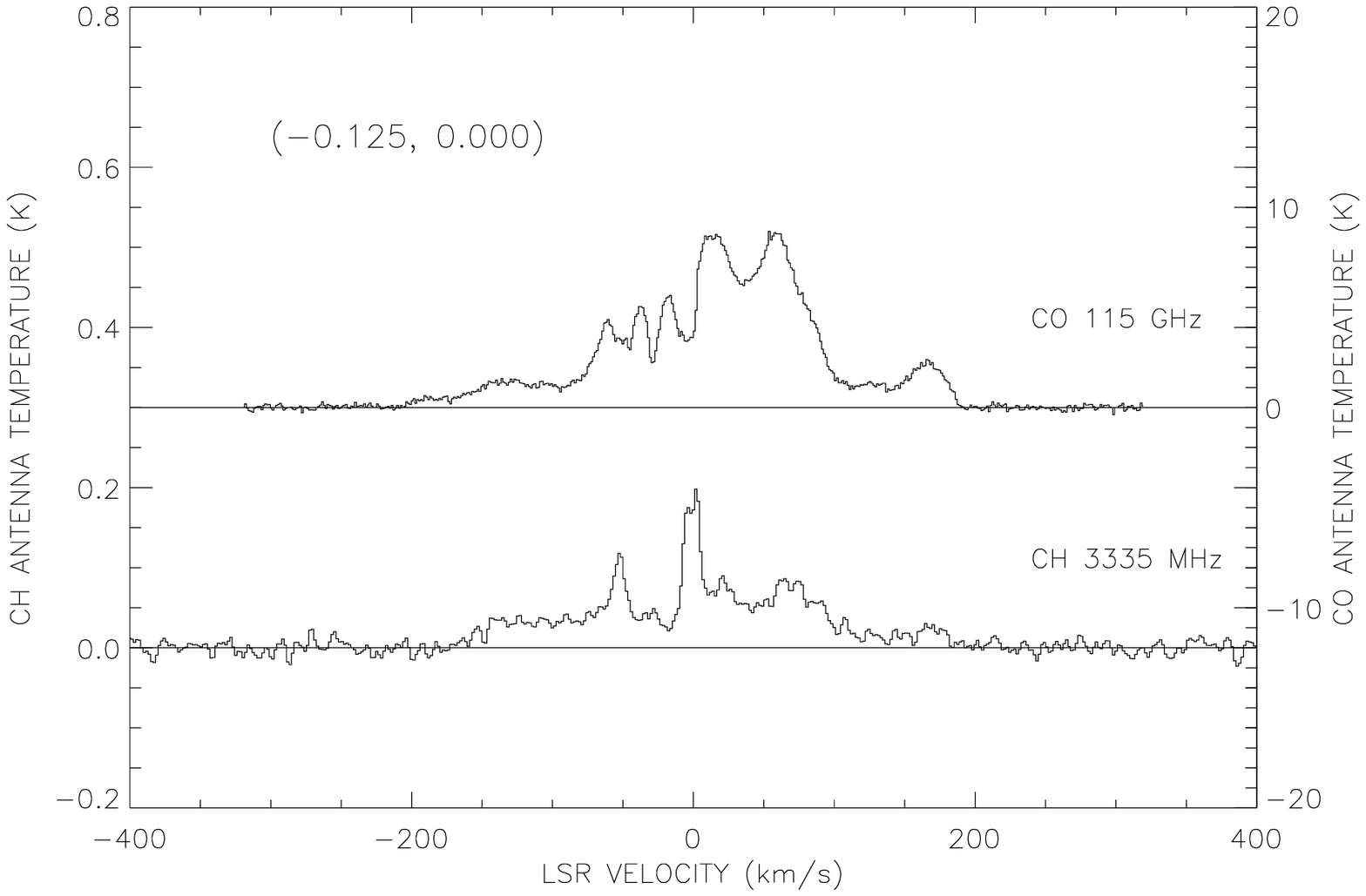}
\caption{
} 
\end{figure}

\begin{figure}
\figurenum{2i}
\plotone{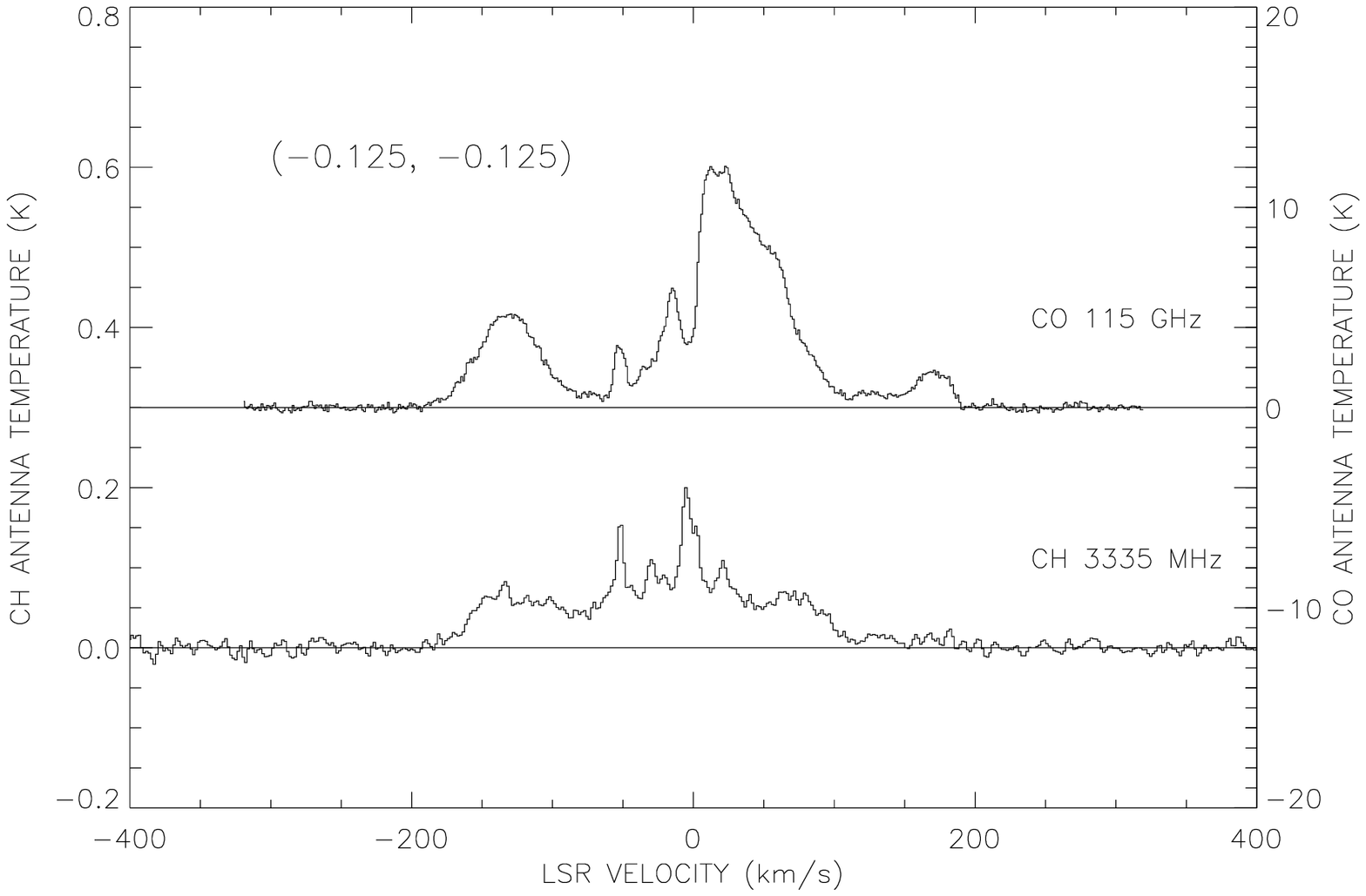}
\caption{
} 
\end{figure}

\begin{figure}
\figurenum{3}
\plotone{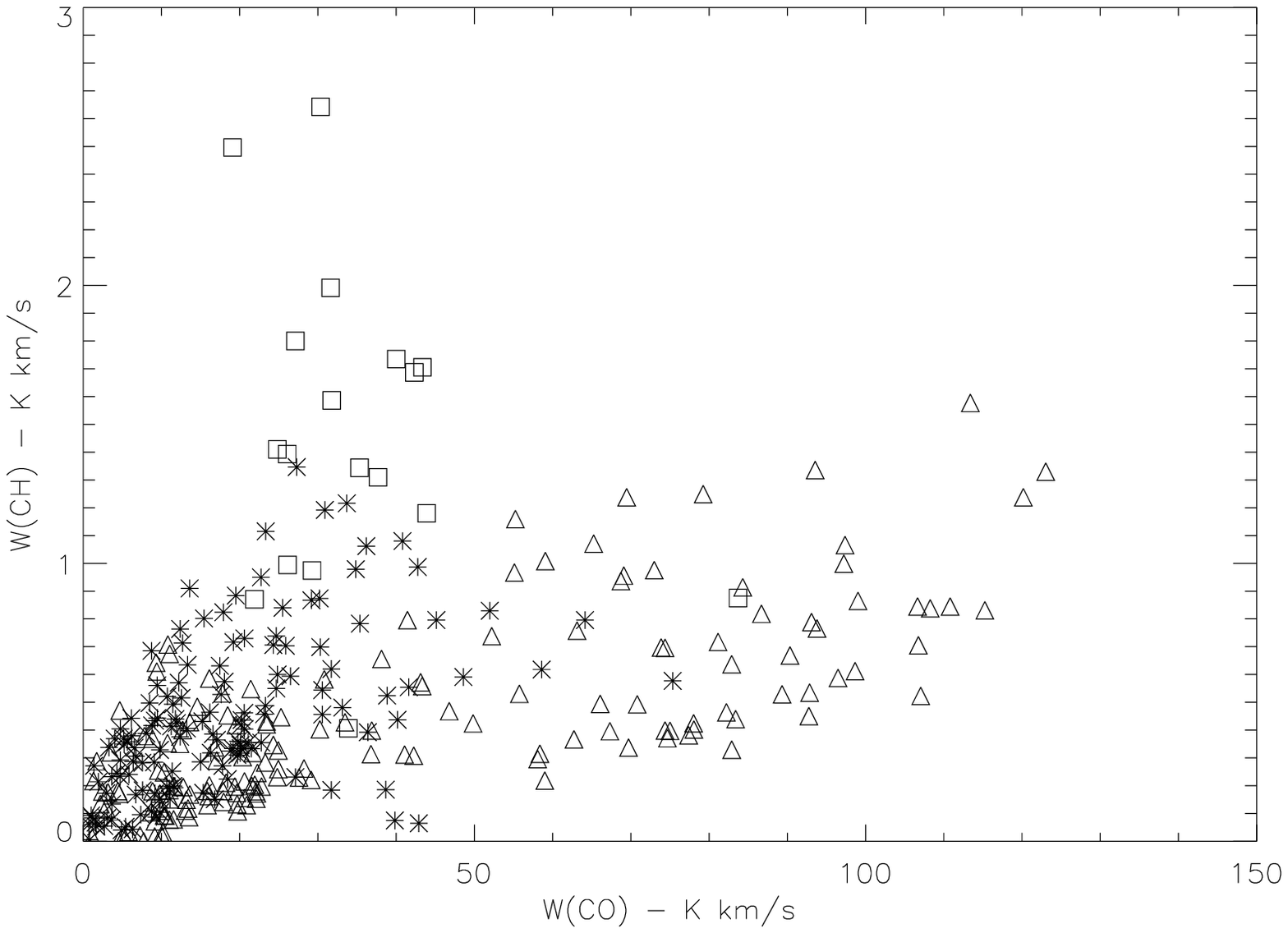}
\caption{
Plot of W$_{CH}$ (the velocity-integrated antenna temperature) versus W$_{CO}$
for all the data in the mapped region binned in velocity intervals of $\sim$ 9 km s$^{-1}$.
The data are broken up into 3 subsets: the triangles represent positive velocity intervals
greater than 9 km s$^{-1}$, the asterisks represent negative velocity intervals less than
$-$9 km s$^{-1}$, and the squares represent the two velocity intervals on either
side of 0 km s$^{-1}$ (see \S 3.1).
}
\label{f3}
\end{figure}

\begin{figure}
\figurenum{4}
\plotone{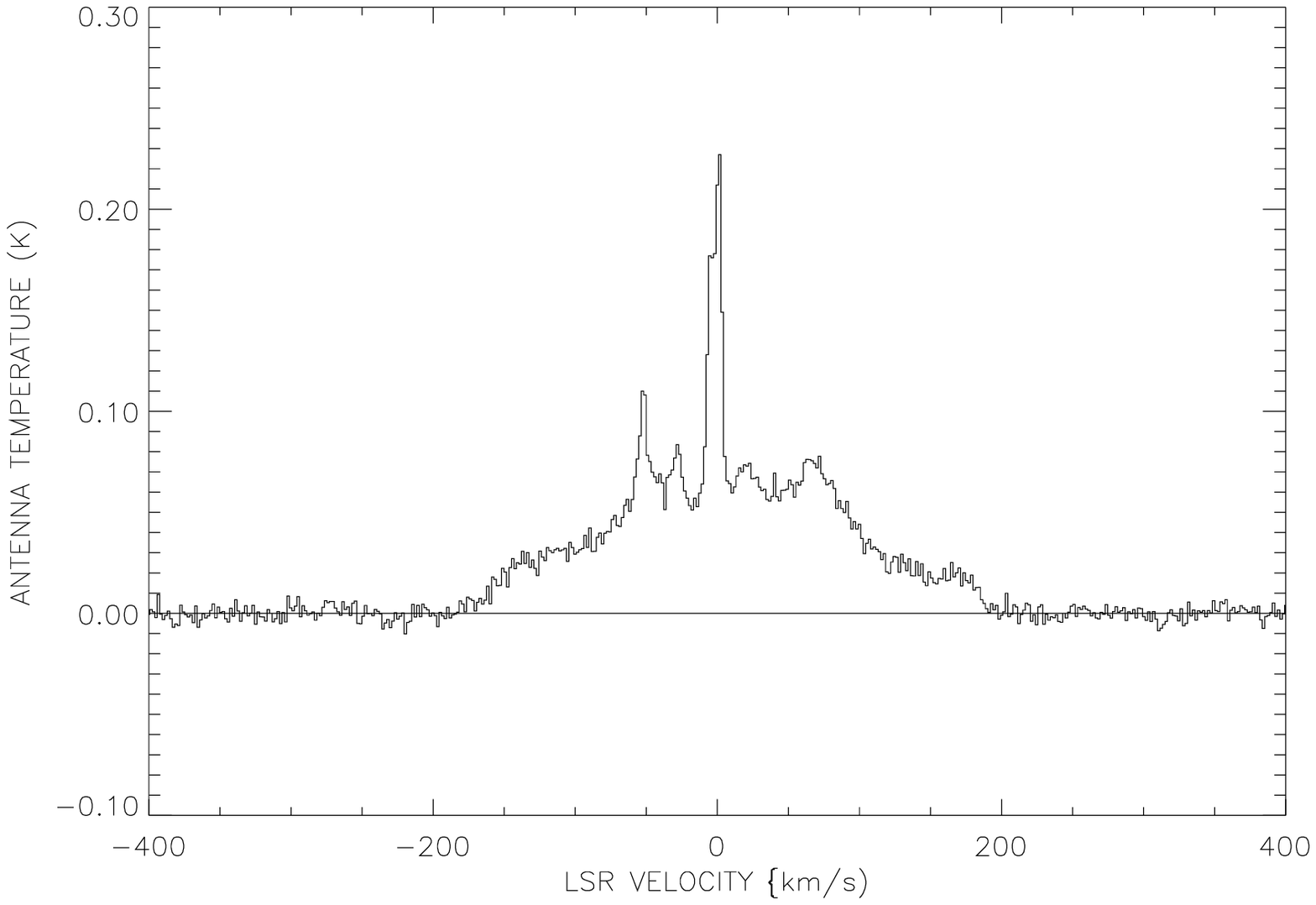}
\caption{
Averaged CH spectrum of the mapped region.  The strong, narrow emission at $\sim$ 0 km s$^{-1}$
is evident and a Gaussian fit to it gives a FWHM 
of 17.6 km s$^{-1}$ centered at v$_{LSR} = -$1.20
km s$^{-1}$ (see \S 4 for details). 
Also noticeable is the emission from the 3-kpc arm at v $\sim -$50 km
s$^{-1}$. 
}
\label{f4}
\end{figure}

\end{document}